%% file: INFOCOM_2009.tex
\documentclass[conference,10pt]{IEEEtran}
\IEEEoverridecommandlockouts

\makeatletter
\def\ps@headings{%
\def\@oddhead{\mbox{}\scriptsize\rightmark \hfil \thepage}%
\def\@evenhead{\scriptsize\thepage \hfil \leftmark\mbox{}}%
\def\@oddfoot{}%
\def\@evenfoot{}}
\makeatother

\usepackage[noadjust]{cite}

\ifCLASSINFOpdf
  \usepackage[pdftex]{graphicx}
  \graphicspath{{./figures/}}
\else
  \usepackage[dvips]{graphicx}
  \graphicspath{{./figures/}}
\fi

\usepackage[cmex10]{amsmath}
\usepackage{amssymb}
\usepackage{amsfonts}
\usepackage{algorithmic}

\usepackage{hyperref}

\input{preamble}

\usepackage{url}

\begin{document}
\title{Network coding meets TCP}
\author{
\IEEEauthorblockN{Jay~Kumar~Sundararajan\authorrefmark{1}, Devavrat~Shah\authorrefmark{1}, Muriel~M\'edard\authorrefmark{1}, Michael Mitzenmacher\authorrefmark{2}, Jo\~ao Barros\authorrefmark{3}}
\IEEEauthorblockA{
\begin{tabular}{ccc}
\ &\ &\ \\
\authorrefmark{1}Dept. of EECS&\authorrefmark{2}School of Eng. and Appl. Sciences&\authorrefmark{3}Dept. of Computer Science\\
Massachusetts Institute of Technology,&Harvard University,&Instituto de Telecomunica\c{c}\~oes\\
Cambridge, MA 02139, USA&Cambridge, MA 02138, USA&Universidade do Porto, Portugal\\
\{jaykumar,devavrat,medard\}@mit.edu&michaelm@eecs.harvard.edu&barros@dcc.fc.up.pt\\
\end{tabular}
}
}
\maketitle

\begin{abstract}
We propose a mechanism that incorporates network coding into TCP with only minor changes to the protocol stack, thereby allowing incremental deployment. In our scheme, the source transmits random linear combinations of packets currently in the congestion window. At the heart of our scheme is a new interpretation of ACKs -- the sink acknowledges every degree of freedom (\ie, a linear combination that reveals one unit of new information) even if it does not reveal an original packet immediately. Such ACKs enable a TCP-like sliding-window approach to network coding. Our scheme has the nice property that packet losses are essentially masked from the congestion control algorithm.  Our algorithm therefore reacts to packet drops in a smooth manner, resulting in a novel and effective approach for congestion control over networks involving lossy links such as wireless links.  Our experiments show that our algorithm achieves higher throughput compared to TCP in the presence of lossy wireless links.  We also establish the soundness and fairness properties of our algorithm. 
\end{abstract}

\section{Introduction}
Network coding has emerged as an important potential approach to the operation of communication networks, especially wireless networks. The major benefit of network coding stems from its ability to {\em mix} data, across time and across flows. This makes data transmission over lossy wireless networks robust and effective. Despite this potential of network coding, we still seem far from seeing widespread implementation of network coding across networks.  We believe a major reason for this is that it is not clear how to naturally add network coding to current network systems (the incremental deployment problem) and how network coding will behave in the wild.

In order to bring the ideas of network coding into practice, we need a protocol that brings out the benefits of network coding while requiring very little change in the protocol stack. Flow control and congestion control in today's internet are predominantly based on the Transmission Control Protocol (TCP), which works using the idea of a sliding transmission window of packets, whose size is controlled based on feedback. The TCP paradigm has clearly proven successful.  We therefore see a need to find a sliding-window approach as similar as possible to TCP for network coding that makes use of acknowledgments for flow and congestion control. (This problem was initially proposed in \cite{desmondfeedback}.) Such an approach would necessarily differ from the generation-based approach more commonly considered for network coding \cite{pracnc}. In this paper, we show how to incorporate network coding into TCP, allowing its use with minimal changes to the protocol stack, and in such a way that incremental deployment is possible.  

The main idea behind TCP is to use acknowledgments of newly received packets as they arrive {\em in correct sequence order} in order to guarantee reliable transport and also as a feedback signal for the congestion control loop. This mechanism requires some modification for systems using network coding. The key difference to be dealt with is that under network coding the receiver does not obtain original packets of the message, but linear combinations of the packets that are then decoded to obtain the original message once enough such combinations have arrived. Hence, the notion of an ordered sequence of packets as used by TCP is missing, and further, a linear combination may bring in new information to a receiver even though it may not reveal an original packet immediately. The current ACK mechanism does not allow the receiver to acknowledge a packet before it has been decoded. For network coding, we need a modification of the standard TCP mechanism that acknowledges every unit of information received. A new unit of information corresponds mathematically to a {\em degree of freedom}; essentially, once $n$ degrees of freedom have been obtained, a message that would have required $n$ unencoded packets can be decoded. We present a mechanism that performs the functions of TCP, namely reliable transport and congestion control, based on acknowledging every degree of freedom received, whether or not it reveals a new packet immediately.

Our solution introduces a new network coding layer between the transport layer and the network layer of the protocol stack. We use the same principle for congestion control as TCP, namely that the number of packets involved in transmissions cannot exceed the number of acknowledgments received by more than the congestion window size. The rules for adapting the congestion window size are also identical to TCP.  However, we introduce two main changes. First, whenever the source is allowed to transmit, it sends a random linear combination of all packets in the congestion window. Second, the receiver acknowledges degrees of freedom and not original packets. (This idea was previously introduced in \cite{ARQforNC} in the context of a single hop erasure broadcast link.) An appropriate interpretation of the degree of freedom allows us to order the receiver degrees of freedom in a manner consistent with the packet order of the transmitter. This lets us utilize the standard TCP protocol with the minimal change. We use the TCP-Vegas protocol, as it is more compatible with our modifications. The rest of the paper explains the details of our new protocol along with its theoretical basis, and analyzes its performance using simulations as well as an idealized theoretical analysis.

In considering the potential benefits of our network coding with a TCP-based protocol, we focus on the area of wireless links.  It is well known that TCP is not well suited for lossy links, which are generally more prevalent in wireless systems. Adapting TCP for wireless scenarios is a very well-studied problem (see \cite{rangwala} and references therein for a survey). Coding across packets is a very natural way to handle losses, and is well-suited to handle the broadcast nature of wireless for a multiple receiver scenario.  Our extension of TCP to a system with coded packets leads to a new approach to implementing TCP over wireless networks, and it is here where the benefits of our approach are most dramatic.

TCP performs poorly on lossy links primarily because it is designed to interpret each loss as a congestion signal. Our new protocol therefore aims to make a lossy channel appear as a lossless channel to TCP, using random linear network coding. Masking losses from TCP has been considered earlier using link layer retransmission \cite{pal95}. However, it has been noted in the literature \cite{dcy93}, \cite{hari} that the interaction between link layer retransmission and TCP's retransmission can be complicated and that performance may suffer due to independent retransmission protocols at different layers. In contrast, our scheme does not rely on the link layers for recovering losses. Instead, we use an erasure correction scheme based on random linear codes between the TCP and IP layers. Our scheme respects the end-to-end philosophy of TCP -- coding operations are performed only at the end hosts.

\subsection{Previous work}
Starting with the initial works of \cite{ahlswede} and \cite{koettermedard}, there has been a rapid growth in the theory and potential applications of network coding. These developments have been summarized in several survey papers and books such as \cite{lunhobook}. However, to a large extent, this theory has not yet been implemented in practical systems.

There have been several important advances in bridging the gap between theory and practice. The distributed random linear coding idea, introduced by Ho \etal. \cite{traceythesis}, is a significant step towards a robust implementation. The work by Chou \etal \cite{pracnc} introduced the idea of embedding the coefficients used in the linear combination in the packet header, and also the notion of generations (coding blocks). The work by Katti \etal \cite{cope} used the idea of local opportunistic coding to present a practical implementation of a network coded system for unicast. 

\section{Preliminaries}\label{sec:prelim}
\noindent We introduce definitions that will be useful throughout the paper (see \cite{ARQforNC} for more details). We treat packets as vectors over a finite field $\F$ of size $q$. All the discussion here is with respect to a single source that generates a stream of packets. The $k^{th}$ packet that the source generates is said to have an \emph{index} $k$ and is denoted as $\mathbf{p_k}$. 

\begin{definition}[Seeing a packet]\label{def:seen}
  A node is said to have \emph{seen} a packet $\mathbf{p_k}$ if it has enough information to compute a linear combination of the form $(\mathbf{p_k} + \mathbf{q})$, where $\mathbf{q} = \sum_{\ell > k} \alpha_\ell \mathbf{p}_\ell$, with $\alpha_\ell \in \F$ for all $\ell > k$. Thus, $\mathbf{q}$ a linear combination involving packets with indices larger than $k$. 
\end{definition}
The notion of ``seeing'' a packet is a natural extension of the notion of ``decoding'' a packet, or more specifically, receiving a packet in the context of classical TCP. For example, if a packet ${\mathbf p_k}$ is decoded then it is indeed also seen, as $\mathbf{q}=\mathbf{0}$. A node can compute any linear combination whose coefficient vector is in the span of the coefficient vectors of previously received linear combinations. This leads to the following definition.

\begin{definition}[Knowledge of a node]
The \emph{knowledge of a node} is the set of all linear combinations of original packets that it can compute, based on the information it has received so far. The coefficient vectors of these linear combinations form a vector space called the \emph{knowledge space} of the node. 
\end{definition}
We state a useful proposition without proof (see Corollary 1, \cite{ARQforNC} for details).
\begin{proposition}\label{witness}
\it If a node has seen packet $\mathbf{p_k}$, then it knows exactly one linear combination of the form $\mathbf{p_k}+\mathbf{q}$ such that $\mathbf{q}$ is itself a linear combination involving only \emph{\textbf{unseen}} packets. 
\end{proposition}
The above proposition inspires the following definition. 
\begin{definition}[Witness]
We call the unique linear combination guaranteed by Proposition \ref{witness}, 
the \emph{witness for seeing $\mathbf{p_k}$}.
\end{definition}

\begin{figure}
  \begin{center}
	\includegraphics[width=0.45\textwidth]{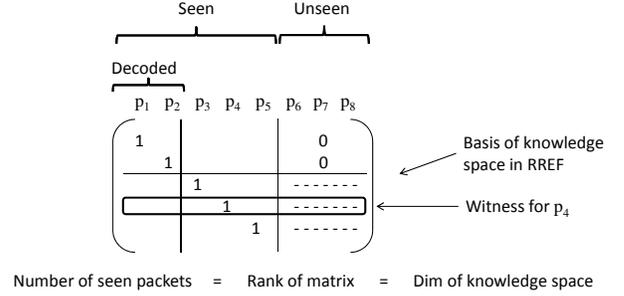}
  \end{center}
  \caption{Seen packets and witnesses in terms of the basis matrix}
  \vspace{-.06in}
  \label{fig:seenpackets}
\end{figure}

A compact representation of the knowledge space is the basis matrix. This is a matrix in row-reduced echelon form (RREF) such that its rows form a basis of the knowledge space. Figure~\ref{fig:seenpackets} explains the notion of a seen packet in terms of the basis matrix. Essentially, the seen packets are the ones that correspond to the pivot columns of the basis matrix. Given a seen packet, the corresponding pivot row gives the coefficient vector for the witness linear combination. An important observation is that \emph{the number of seen packets is always equal to the dimension of the knowledge space}, or the number of degrees of freedom that have been received so far. A newly received linear combination that increases the dimension is said to be \emph{innovative}. We assume throughout the paper that the field size is very large. As a consequence, each reception will be innovative with high probability, and will cause the next unseen packet to be seen (see Lemma \ref{lemma:seen}). 

{\it Example:} Suppose a node knows the following linear combinations: $\mathbf{x} = (\mathbf{p_1}+\mathbf{p_2})$ and $\mathbf{y}=(\mathbf{p_1}+\mathbf{p_3})$. Since these are linearly independent, the knowledge space has a dimension of 2. Hence, the number of seen packets must be 2. It is clear that packet $\mathbf{p_1}$ has been seen, since $\mathbf{x}$ satisfies the requirement of Definition \ref{def:seen}. Now, the node can compute $\mathbf{z}\triangleq \mathbf{x}-\mathbf{y}=(\mathbf{p_2}-\mathbf{p_3})$. Thus, it has also seen $\mathbf{p_2}$. That means $\mathbf{p_3}$ is unseen. Hence, $\mathbf{y}$ is the witness for $\mathbf{p_1}$, and $\mathbf{z}$ is the witness for $\mathbf{p_2}$.
\section{The new protocol}\label{sec:protocol}
In this section, we present the logical description of our new protocol, followed by a way to implement these ideas with as little disturbance as possible to the existing protocol stack.

\subsection{Logical description}
The main aim of our algorithm is to mask losses from TCP using random linear coding. We make some important modifications in order to incorporate coding. First, instead of the original packets, we transmit random linear combinations of packets in the congestion window. 
While such coding helps with erasure correction, it also leads to a problem in acknowledging data. TCP operates with units of packets, which have a well-defined ordering. Thus, the packet sequence number can be used for acknowledging the received data. The unit in our protocol is a degree of freedom. However, when packets are coded together, there is no clear ordering of the degrees of freedom that can be used for ACKs. Our main contribution is the solution to this problem. The notion of seen packets defines an ordering of the degrees of freedom that is consistent with the packet sequence numbers, and can therefore be used to acknowledge degrees of freedom. 

Upon receiving a linear combination, the sink finds out which packet, if any, has been newly seen because of the new arrival and acknowledges that packet. The sink thus pretends to have received the packet even if it cannot be decoded yet. We will show in Section \ref{sec:soundness} that at the end this is not a problem because if all the packets in a file have been seen, then they can all be decoded as well.

The idea of transmitting random linear combinations and acknowledging seen packets achieves our goal of masking losses from TCP as follows. As mentioned in Section \ref{sec:prelim}, with a large field size, every random linear combination is very likely to cause the next unseen packet to be seen in order. So, even if a transmitted linear combination is lost, the next unseen packet will eventually be seen by the receiver in the form of the next linear combination that is successfully received. {F}rom TCP's perspective, this appears as though the degree of freedom waits in a fictitious queue until the channel stops erasing packets and allows it through. Thus, there will never be any duplicate ACKs. Every ACK will cause the congestion window to advance. In short, \emph{the lossiness of the link is presented to TCP as an additional queuing delay that leads to a larger effective round-trip time}. The more lossy the link is, the larger will be the RTT that TCP sees.

The natural question that arises is -- how does this affect congestion control? Since we mask losses from the congestion control algorithm, the TCP-Reno style approach to congestion control using packet loss as a congestion indicator is not well suited to this  situation. However, it is useful to note that the congestion related losses are also made to appear as a longer RTT. Therefore, we need an approach that infers congestion from an increase in RTT. The natural choice is TCP-Vegas. 

TCP-Vegas uses a proactive approach to congestion control by inferring the size of the network buffers even before they start dropping packets. The crux of the algorithm is to estimate the round-trip time (RTT) and use this information to find the discrepancy between the expected and actual transmission rate. As congestion arises, buffers start to fill up and the RTT starts to rise, and this is used as the congestion signal. This signal is used to adjust the congestion window and hence the rate. For further details, the reader is referred to \cite{tcpvegas}. 

In order to use TCP-Vegas correctly in this setting, we need to feed it the fictitiously longer RTT of a degree of freedom that includes the fictitious queuing delay. We introduce a novel RTT estimation algorithm to do this. 

The sender can note down the transmission time of every linear combination. So the question is, when an ACK arrives, to which transmission should it be matched in order to compute the RTT? Our solution is to match it to the transmission that occurred after the one that triggered the previous ACK.

\begin{figure}
  \begin{center}
	\includegraphics[width=0.4\textwidth]{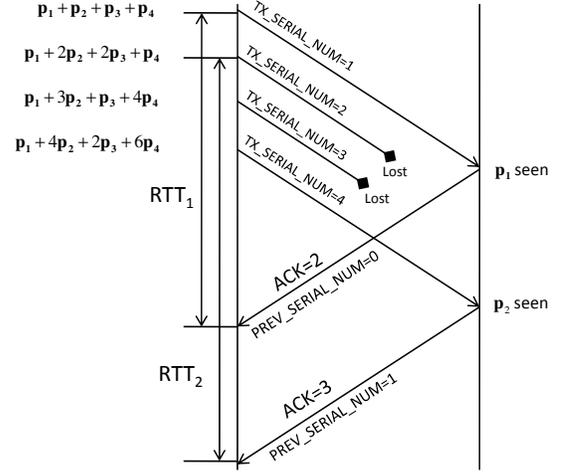}
  \end{center}
	\caption{Example of coding, ACK and RTT measurement}
  \vspace{-.06in}
  \label{fig:rttexample}
\end{figure}

Consider the example shown in Figure \ref{fig:rttexample}. The congestion window is assumed to be 4 packets long. All 4 transmissions are linear combinations of the 4 packets in the window. 
In this example, the $1^{st}$ packet is seen because of the $1^{st}$ transmission. The $2^{nd}$ and $3^{rd}$ transmissions are lost, and the $4^{th}$ transmission causes the $2^{nd}$ packet to be seen (the discrepancy is because of losses). As far as the RTT estimation is concerned, transmissions 2, 3 and 4 are treated as attempts to convey the $2^{nd}$ degree of freedom. The RTT for the $2^{nd}$ packet is therefore computed based on the oldest such attempt, namely the $2^{nd}$ transmission. In other words, the RTT is the difference between the time of reception of ACK=3 (in the figure), and the time of the transmission of $(\mathbf{p_1}+2\mathbf{p_2}+2\mathbf{p_3}+\mathbf{p_4})$. The implementation of this idea is explained in the next subsection.

\subsection{Implementation}
\begin{figure}
  \begin{center}
	\includegraphics[width=0.4\textwidth]{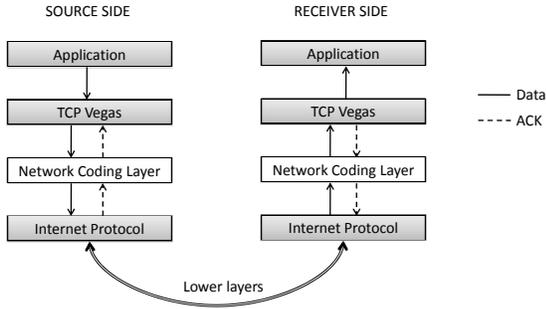}
  \end{center}
  \caption{New network coding layer in the protocol stack}
  \vspace{-.06in}
  \label{fig:layers}
\end{figure}

The implementation of all these ideas in the existing protocol stack needs to be done in as non-intrusive a manner as possible. We present a solution which embeds the network coding operations in a separate layer below TCP and above IP on the source and receiver side, as shown in Figure \ref{fig:layers}. The exact operation of these modules is described next.

The sender module accepts packets from the TCP source and buffers them into an encoding buffer which represents the coding window\footnote{Whenever a new packet enters the TCP congestion window, TCP transmits it to the network coding module, which then includes it in the coding window. Thus, the coding window is related to the TCP layer's congestion window. However, it is generally not identical to the congestion window. In particular, the coding window will still include packets that were transmitted earlier by TCP, but are no longer in the congestion window because of a reduction of the window size by TCP. However, this is not a problem because including more packets in the linear combination will only increase its chances of being innovative.}, until they are ACKed by the receiver. The sender then generates and sends random linear combinations of the packets in the coding window. The coefficients used in the linear combination are also conveyed in the header. 

For every packet that arrives from TCP, $R$ linear combinations are sent to IP on average, where $R$ is the redundancy parameter. The average rate at which linear combinations are sent into the network is thus a constant factor more than the rate at which TCP's congestion window progresses. This is necessary in order to compensate for the loss rate of the channel and to match TCP's sending rate to the rate at which data is actually sent to the receiver. If there is too little redundancy, then the data rate reaching the receiver will not match the sending rate because of the losses. This leads to a situation where the losses are not effectively masked from the TCP layer. Hence, there are frequent timeouts leading to a low throughput. On the other extreme, too much redundancy is also bad, since then the transmission rate becomes limited by the rate of the code itself. Besides, sending too many linear combinations can congest the network. The ideal level of redundancy is to keep $R$ equal to the reciprocal of the probability of successful reception. Thus, in practice the value of $R$ should be dynamically adjusted by estimating the loss rate, possibly using the RTT estimates. 

Upon receiving a linear combination, the receiver module first retrieves the coding coefficients from the header and appends it to the basis matrix of its knowledge space. Then, it performs a Gaussian elimination to find out which packet is newly seen so that this packet can be ACKed. The receiver module also maintains a buffer of linear combinations of packets that have not been decoded yet. Upon decoding the packets, the receiver module delivers them to the TCP sink. 

We now describe the implementation of the RTT computation. As in TCP-Vegas, the sender module notes down the system clock corresponding to every transmission. In addition, the transmitter embeds in the header of every transmission a transmit serial number $TX\_SERIAL\_NUM$. This serial number is used for identifying the transmissions while computing RTT. Now, in every ACK, the sink embeds in the header \emph{the transmit serial number of that packet, the reception of which, triggered the sink's previous ACK}. This is called $PREV\_SERIAL\_NUM$. (See the example in Figure \ref{fig:rttexample}.) Upon receiving the ACK, the transmitter first notes the sequence number of the packet that is being ACKed. Then, for purposes of RTT computation, it matches this ACK to the transmission whose serial number is $(PREV\_SERIAL\_NUM+1)$. The transmit time of the matched transmission is then loaded into the transmit timestamp echo field of the TCP ACK packet and delivered to TCP. Thus, the TCP sender is intentionally mislead into computing the fictitious RTT for the degree of freedom.
	
The algorithm is specified below using pseudo-code. This specification assumes a one-way TCP with the timestamps option turned on.
\subsubsection{Source side}
The source side algorithm has to respond to two types of events -- the arrival of a packet from the source TCP, and the arrival of an ACK from the receiver via IP. 
\begin{enumerate}
  \item Set $TX\_SERIAL\_NUM$ and $NUM$ to 0.
  \item {\it Wait state:} If any of the following events occurs, respond as follows; else, wait.
  \item {\it Packet arrives from TCP:} 
	\begin{enumerate}
	  \item If the packet is a control packet used for connection management, deliver it to the IP layer and return to wait state.
	  \item If packet is not already in the coding window, add it to the coding window. 
	  \item Set $NUM:=NUM+R$. ($R$ is the redundancy factor.)
	  \item Repeat the following $\lfloor NUM \rfloor$ times:
	  
		  i) Increment $TX\_SERIAL\_NUM$ by 1.
		  
		  ii) Generate a random linear combination of the packets in the coding window. 
		  
		  iii) Add the network coding layer header to it that contains the following: the coefficients used for the random linear combination in terms of the packets in the current coding window, the set of packets in the window, and $TX\_SERIAL\_NUM$. 
		  
		  iv) Deliver the packet to the IP layer.
		  
		  v) Note down the current time as the transmission time corresponding to $TX\_SERIAL\_NUM$.
		  
	  \item Set $NUM:=$ fractional part of $NUM$.
	  \item Return to the wait state.
	\end{enumerate}
  \item {\it ACK arrives from receiver:}
	\begin{enumerate}
	  \item Remove the network coding ACK header and retrieve $PREV\_SERIAL\_NUM$. 
	  \item Modify the TCP ACK header as follows. Set the timestamp-echo field in the header to the transmission time corresponding to the transmission number $(PREV\_SERIAL\_NUM + 1)$. 
	\end{enumerate}
\end{enumerate}

\subsubsection{Receiver side}
On the receiver side, the algorithm again has to respond to two types of events: the arrival of a packet from the source, and the arrival of ACKs from the TCP sink.

\begin{enumerate}
  \item {\it Wait state: } If any of the following events occurs, respond as follows; else, wait.  
  \item {\it ACK arrives from TCP sink:} 
	If the ACK is a control packet for connection management, deliver it to the IP layer and return to the wait state; else, ignore the ACK.
  \item {\it Packet arrives from source side:}
	\begin{enumerate}
	  \item Remove the network coding header and retrieve the coding vector as well as the $TX\_SERIAL\_NUM$. 
	  \item Add the coding vector as a new row to the existing coding coefficient matrix, and perform Gaussian elimination to update the set of seen packets.
	  \item Add the payload to the decoding buffer. Perform the operations corresponding to the Gaussian elimination, on the buffer contents. If any packet gets decoded in the process, deliver it to the TCP sink.
	  \item Generate a new TCP ACK with sequence number equal to that of the oldest unseen packet. 
	  \item Add the network coding ACK header to the ACK, consisting of the current value of $PREV\_SERIAL\_NUM$. 
	  \item Update $PREV\_SERIAL\_NUM$ to the $TX\_SERIAL\_NUM$ of the new arrival.
	\end{enumerate}
\end{enumerate}
\section{Soundness of the protocol}\label{sec:soundness}
We argue that our protocol guarantees reliable transfer of information. In other words, every packet in the packet stream generated by the application at the source will be delivered eventually to the application at the sink. We observe that the acknowledgment mechanism ensures that the coding module at the sender does not remove a packet from the coding window unless it has been ACKed, \ie, unless it has been seen by the sink. Thus, we only need to argue that if all packets in a file have been seen, then the file can be decoded at the sink. 

\begin{theorem}\label{thm:soundness}
From a file of $n$ packets, if every packet has been seen, then every packet can also be decoded.
\end{theorem}
\IEEEproof
If the sender knows a file of $n$ packets, then the sender's knowledge space is of dimension $n$.  Every seen packet corresponds to a new dimension. Hence, if all $n$ packets have been seen, then the receiver's knowledge space is also of dimension $n$, in which case it must be the same as the sender's and all packets can be decoded.
\endproof
In other words, seeing $n$ different packets corresponds to having $n$ linearly independent equations in $n$ unknowns. Hence, the unknowns can be found by solving the system of equations. At this point, the file can be delivered to the TCP sink. In practice, one does not have to necessarily wait until the end of the file to decode all packets. Some of the unknowns can be found even along the way. In particular, whenever the number of equations received catches up with the number of unknowns involved, the unknowns can be found. Now, for every new equation received, the receiver sends an ACK. The congestion control algorithm uses the ACKs to control the injection of new unknowns into the coding window. Thus, the discrepancy between the number of equations and number of unknowns does not tend to grow with time, and therefore will hit zero often based on the channel conditions. As a consequence, the decoding buffer will tend to be stable.

An interesting observation is that the arguments used to show the soundness of our approach are quite general and can be extended to more general scenarios such as random linear coding based multicast over arbitrary topologies.

\section{Fairness of the protocol}
Here, we study the fairness property of our algorithm through simulations. 
\subsection{Simulation setup}\label{sec:simsetup}
\begin{figure}
  \begin{center}
	\includegraphics[width=0.43\textwidth]{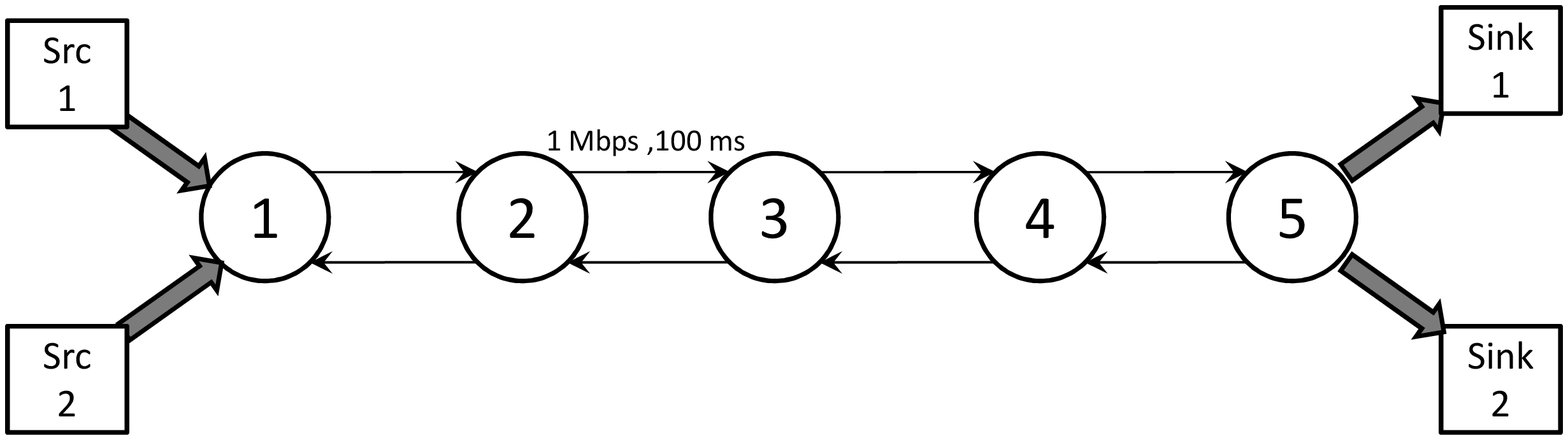}
  \end{center}
  \caption{Simulation topology}
  \vspace{-.06in}
  \label{fig:simtop}
\end{figure}

\noindent The protocol described above is simulated using the Network Simulator (ns-2) \cite{ns2}. The topology for all the simulations is a tandem network consisting of 4 hops (hence 5 nodes), shown in Figure \ref{fig:simtop}. The source and sink nodes are at opposite ends of the chain. Two FTP applications want to communicate from the source to the sink. They either use TCP without coding or TCP with network coding (denoted TCP/NC). All the links have a bandwidth of 1 Mbps, and a propagation delay of 100 \emph{ms}. The buffer size on the links is set at 200. The TCP receive window size is set at 100 packets, and the packet size is 1000 bytes. The Vegas parameters are chosen to be $\alpha=28, \beta=30, \gamma=2$ (see \cite{tcpvegas} for details of Vegas).

\subsection{Fairness and compatibility -- simulation results}
By fairness, we mean that if two similar flows compete for the same link, they must receive an approximately equal share of the link bandwidth. In addition, this must not depend on the order in which the flows join the network. The fairness of TCP-Vegas is a well-studied problem. It is known that depending on the values chosen for the $\alpha$ and $\beta$ parameters, TCP-Vegas could be unfair to an existing connection when a new connection enters the bottleneck link (\cite{vegasfairness}, \cite{vegasnote}). Several solutions have been presented to this problem in the literature (for example, see \cite{renovegas} and references therein). In our simulations, we first pick values of $\alpha$ and $\beta$ that allow fair sharing of bandwidth when two TCP flows without our modification compete with each other, in order to evaluate the effect of our modification on fairness. 

Then, with the same $\alpha$ and $\beta$, we consider two cases:

{\it Case 1:} The situation where two network coded TCP flows compete with each other. 

{\it Case 2:} The situation where a coded TCP flow competes with another flow running TCP without coding. 

\begin{figure}
  \begin{center}
	\includegraphics[width=0.43\textwidth]{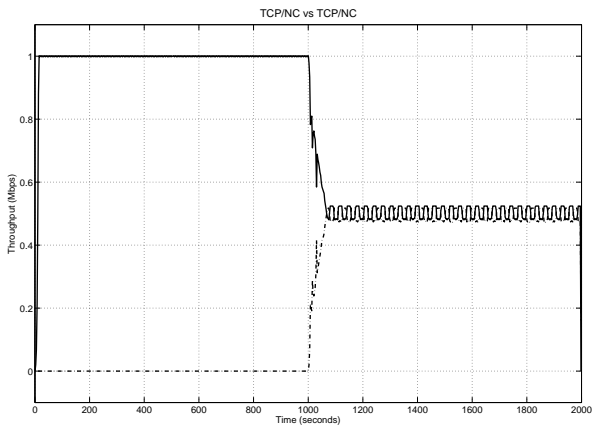}
  \end{center}
  \caption{Fairness - two TCP/NC flows}
  \vspace{-0.1in}
  \label{fig:fair1}
\end{figure}

In both cases, the loss rate is set to 0\% and the redundancy parameter is set to 1 for a fair comparison. In the first simulation, where both flows use TCP/NC, one flow is started at $t=0.5 s$ and the other flow is started at $t=1000 s$. The system is simulated for 2000 $s$. The current throughput is calculated at intervals of $2.5 s$. The evolution of the throughput over time is shown in Figure \ref{fig:fair1}. The figure shows that the effect of introducing the coding layer does not affect fairness. We see that after the second flow starts, the bandwidth gets redistributed fairly. 

\begin{figure}
  \begin{center}
	\includegraphics[width=0.43\textwidth]{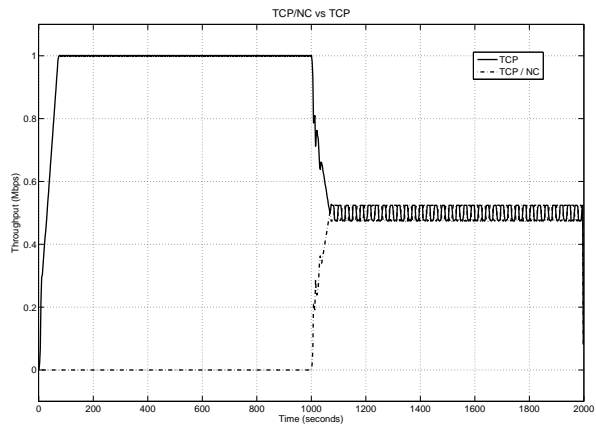}
  \end{center}
  \caption{Fairness and compatibility - one TCP/NC and one TCP flow}
  \vspace{-.1in}
  \label{fig:fair2}
\end{figure}

For case 2, the experiment is repeated, but this time with the TCP flow starting first, and the TCP/NC flow starting at $1000 s$. The corresponding plot is shown in Figure \ref{fig:fair2}. This figure shows that coding is compatible with TCP in the absence of losses. Again we see that after the new flow joins, the bandwidth is divided fairly between the two flows.

\section{Effectiveness of the protocol}
We now show that the new protocol indeed achieves a high throughput, especially in the presence of losses. We first describe simulation results comparing the protocol's performance with that of TCP in Section \ref{sec:tputsim}. Next, in Section \ref{sec:ideal}, we study the effectiveness of the random linear coding ideas in a theoretical model with idealized assumptions such as infinite buffer space, and known channel capacity. We show that in such a scenario, our scheme stabilizes the queues for all rates below capacity. 

\subsection{Throughput of the new protocol -- simulation results}\label{sec:tputsim}
The simulation setup is identical to that used in the fairness simulations (see Section \ref{sec:simsetup}).

We first study the effect of the redundancy parameter on the throughput of TCP/NC for a fixed loss rate of 5\%. By loss rate, we mean the probability of a packet getting lost on each link. Both packets in the forward direction as well as ACKs in the reverse direction are subject to these losses. Since no re-encoding is allowed at the intermediate nodes, the overall probability of packet loss across 4 hops is given by $1-(1-0.05)^4$ which is roughly 19\%. Hence the capacity is roughly 0.81 Mbps, which when split fairly gives 0.405 Mbps per flow. The simulation time is $10000 s$. 

\begin{figure}
  \begin{center}
	\includegraphics[width=0.43\textwidth]{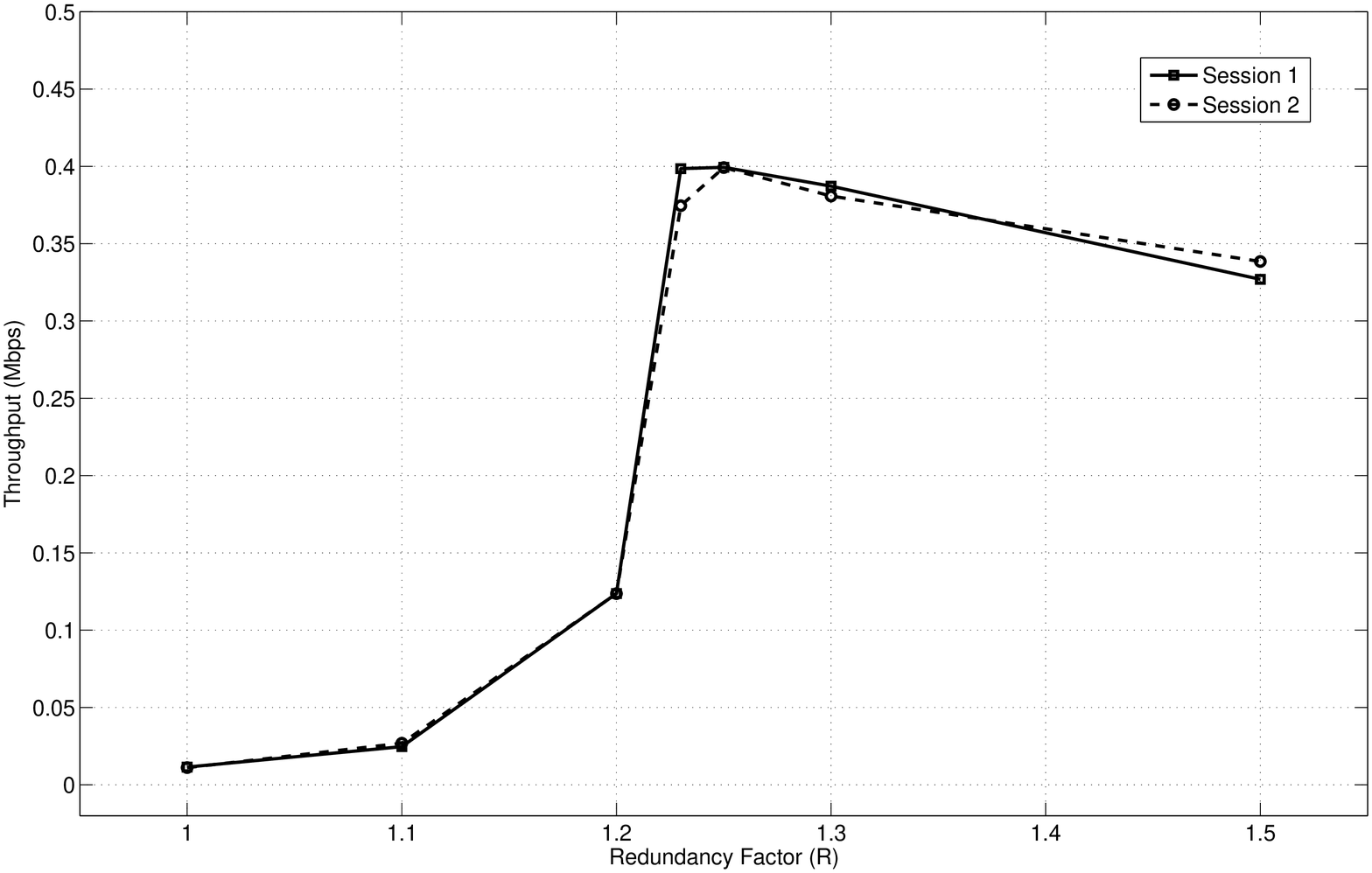}
  \end{center}
  \caption{Throughput vs redundancy for TCP/NC}
  \vspace{-.06in}
  \label{fig:redund}
\end{figure}

We allow two TCP/NC flows to compete on this network, both starting at $0.5 s$. Their redundancy parameter is varied between 1 and 1.5. The theoretically optimum value is approximately $1/(1-0.19) \simeq 1.23$. Figure \ref{fig:redund} shows the plot of the throughput for the two flows, as a function of the redundancy parameter $R$. It is clear from the plot that $R$ plays an important role in TCP/NC. We can see that the throughput peaks around $R=1.25$. The peak throughput achieved is 0.399 Mbps, which is indeed close to the capacity that we calculated above. In the same situation, when two TCP flows compete for the network, the two flows see a throughput of 0.0062 and 0.0072 Mbps respectively. Thus, with the correct choice of $R$, the throughput for the flows in the TCP/NC case is very high compared to the TCP case. In fact, even with $R=0$, TCP/NC achieves about 0.011 Mbps for each flow improving on TCP by almost a factor of 2. 

\begin{figure}
  \begin{center}
	\includegraphics[width=0.43\textwidth]{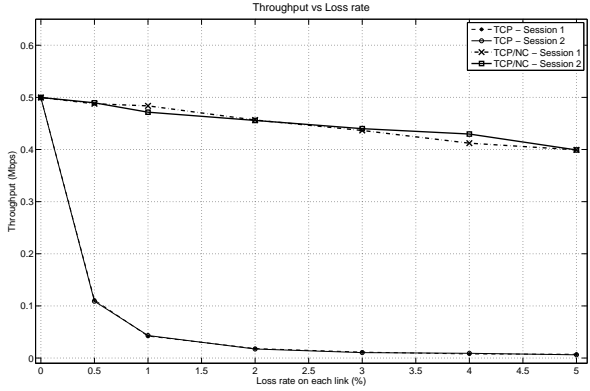}
  \end{center}
  \caption{Throughput vs loss rate for TCP and TCP/NC}
  \vspace{-.06in}
  \label{fig:lossrate}
\end{figure}

Next, we study the variation of throughput with loss rate for both TCP and TCP/NS. The simulation parameters are all the same as above. The loss rate of all links is kept at the same value, and this is varied from 0 to 5\%. We compare two scenarios -- two TCP flows competing with each other, and two TCP/NC flows competing with each other. For the TCP/NC case, we set the redundancy parameter at the optimum value corresponding to each loss rate. Figure \ref{fig:lossrate} shows that TCP's throughput falls rapidly as losses increase. However, TCP/NC is very robust to losses and reaches a throughput that is close to capacity.

\begin{remark}
These simulations are meant to be a preliminary study of our algorithm's performance. They do not account for the overhead associated with the network coding headers while computing the throughput. The main overhead is in conveying the coding coefficients and the contents of the coding window. However, if the source and sink share a pseudorandom number generator, then the coding coefficients can be conveyed by simply sending the current state of the generator. Similarly, the coding window contents can be conveyed in an incremental manner to reduce the overhead. 

Another source of throughput loss that has not been modeled in the simulation is the field size not being large enough. This could cause received linear combinations to be either non-innovative, or might cause packets to be seen out of order, resulting in duplicate ACKs. However, the probability that such problems persist for a long time falls rapidly with the field size. We believe that with practical choices of field size, these issues with only cause transient effects that will not have a significant impact on performance. The exact quantification of these effects remains to be done.
\end{remark}

\subsection{The ideal case}\label{sec:ideal}
In this section, we focus on an idealized scenario in order to provide a first order analysis of our new protocol. We aim to explain the key ideas of our protocol with emphasis on the interaction between the coding operation and the feedback. The model used in this section will also serve as a platform which we can build on to incorporate more practical situations.

We abstract out the congestion control aspect of the problem by assuming that the capacity of the system is fixed in time and known at the source, and hence the arrival rate is always maintained below the capacity. We also assume that nodes have infinite capacity buffers to store packets. We focus on a topology that consists of a chain of erasure-prone links in tandem, with perfect end-to-end feedback from the sink directly to the source. In such a system, we investigate the behavior of the queue sizes at various nodes.

\subsubsection{System model}
\begin{figure}
  \begin{center}
	\includegraphics[width=0.3\textwidth]{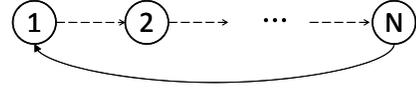}
  \end{center}
  \caption{Topology: Daisy chain with perfect end-to-end feedback}
  \vspace{-.1in}
  \label{fig:daisychain}
\end{figure}

The network we study in this section is a daisy chain of $N$ nodes, each node being connected to the next one by a packet erasure channel, as shown in Figure \ref{fig:daisychain}. We assume a slotted time system. The source generates packets according to a Bernoulli process of rate $\lambda$ packets per slot. The point of transmission is at the very beginning of a slot. Just after this point, every node transmits one random linear combination of the packets in its queue. We ignore propagation delay. Thus, the transmission, if not erased by the channel, reaches the next node in the chain almost immediately. However, the node may use the newly received packet only in the next slot's transmission. We assume perfect, delay-free feedback from the sink to the source. In every slot, the sink generates the feedback signal after the instant of reception of the previous node's transmission. The erasure event happens with a probability $(1-\mu_i)$ on the channel connecting node $i$ and $(i+1)$, and is assumed to be independent across different channels and over time. Thus, the system has a capacity $\min_i\mu_i$ packets per slot. We assume that $\lambda<\min_i\mu_i$, and define the load factor $\rho_i=\lambda/\mu_i$. The relation between the transmitted linear combination and the original packet stream is conveyed in the packet header. We ignore this overhead for the analysis in this section.
\begin{remark}\label{rem:noncoding}
This model and the following analysis also works for the case when not all intermediate nodes are involved in the network coding. If some node simply forwards the incoming packets, then we can incorporate this in the following way. An erasure event on either the link entering this node or the link leaving this node will cause a packet erasure. Hence, these two links can be replaced by a single link whose probability of being ON is simply the product of the ON probabilities of the two links being replaced. Thus, all non-coding nodes can be removed from the model, which brings us back to the same situation as in the above model.
\end{remark}

\subsubsection{Queue update mechanism}
As specified in the previous subsection, the coding scheme we consider is one where each node transmits a random linear combination of the current contents of its queue. Therefore, the question of how to update the queue contents becomes important. In our scheme, the sink sends an ACK to the source in every slot, using the feedback link. The ACK contains the index of the oldest packet not yet seen by the sink. Upon receiving the ACK, the source drops all packets from its queue with an index lower than the sink's request. 
As for the intermediate nodes, they do not have direct feedback from the sink. Therefore, the source has to inform them about the sink's ACK. This information is sent on the same packet erasure channel used for the regular transmission. This feed-forward of the sink's status is modeled in our setup as follows. Whenever the channel entering an intermediate node is in the ON state (\ie, no erasure), the node's version of the sink's status is updated to that of the previous node. In practice, the source need not transmit the sink's status explicitly. The intermediate nodes can infer it from the set of packets that have been involved in the linear combination -- if a packet is no longer involved, that means the source must have dropped it, implying that the sink must have ACKed it already. 
Whenever an intermediate node receives an innovative packet, this causes the node to see a previously unseen packet. The node performs a Gaussian elimination to compute the witness of the newly seen packet, and adds this to the queue. Thus, intermediate nodes store the witnesses of the packets that they have seen. The queue update rule is similar to that of the source. An intermediate node drops the witness of all packets up to but excluding the one requested by the sink. This is based on the most updated version of the sink's status known at the intermediate node.

\subsubsection{Queuing analysis}
The following theorem shows that if we allow coding at intermediate nodes, then it is possible to achieve the capacity of the network, namely $\min_i \mu_i$. Note that this theorem also implies that if we only allow forwarding at some of the intermediate nodes, then we can still achieve the capacity of a new network derived by collapsing the links across the non-coding nodes, as described in Remark \ref{rem:noncoding}. 
\begin{theorem}\label{thm:queue}
As long as $\lambda<\mu_k$ for all $0 \le k<N$, the queues at all the nodes will be stable. The expected queue size in steady state at node $k$ ($0 \le k<N$) is given by:
\[\mathbb{E}[Q_k]=\sum_{i=k}^{N-1} \frac{\rho_i(1-\mu_i)}{(1-\rho_i)}+\sum_{i=1}^{k-1}\rho_i\]
\end{theorem}
{\it An implication:} Consider a case where all the $\rho_i$'s are equal to some $\rho$. Then, the above relation implies that in the limit of heavy traffic, \ie, $\rho\rightarrow 1$, the queues are expected to be longer at nodes near the source than near the sink. 

\noindent{\it A useful lemma:}

The following lemma shows that the random linear coding scheme has the property that every time there is a successful reception at a node, the node sees the next unseen packet with high probability, provided the field is large enough. This fact will prove useful while analyzing the evolution of the queues.
\begin{lemma}\label{lemma:seen}
  Let $S_A$ and $S_B$ be the set of packets seen by two nodes A and B respectively. Assume $S_A\backslash S_B$ is non-empty. Suppose A sends a random linear combination of its witnesses of packets in $S_A$ and B receives it successfully. The probability that this transmission causes B to see the oldest packet in $S_A\backslash S_B$ is $\left(1-\frac 1q\right)$, where $q$ is the field size.
\end{lemma}

\IEEEproof 
Let $M_A$ be the basis matrix of A's knowledge space. Then, the coefficient vector of the transmitted linear combination is given by $\mathbf{t}=\mathbf{u}M_A$, where $\mathbf{u}$ is a vector of length $|S_A|=m$ whose entries are independent and uniformly distributed over the finite field $\F$. Now, the entries of $\mathbf{t}$ corresponding to the packets seen by A (\ie, pivot columns of A) are equal to the entries of $\mathbf{u}$ that multiplied the corresponding pivot rows. Therefore, any entry of $\mathbf{t}$ corresponding to a packet seen by A is uniformly distributed and is independent of all other entries of $\mathbf{t}$.

Let $M_B$ be the basis matrix of B's knowledge space before the new reception. Let $d^*$ denote the index of the oldest packet in $S_A\backslash S_B$. Suppose $\mathbf{t}$ is successfully received by B. Then, B will append $\mathbf{t}$ as a new row to $M_B$ and perform Gaussian elimination. The first step involves subtracting from $\mathbf{t}$, suitably scaled versions of the pivot rows such that all entries of $\mathbf{t}$ corresponding to pivot columns of $M_B$ become 0.  (This is possible only if the received packet is innovative.) We need to find the probability that after this step, the leading non-zero entry occurs in column $d^*$, which corresponds to the event that B sees packet $d^*$. Subsequent steps in the Gaussian elimination will not affect this event. Hence, we focus on the first step.

Let $P_B$ denote the set of indices of pivot columns of $M_B$. In the first step, the entry in column $d^*$ of $\mathbf{t}$ becomes \[t'(d^*)=t(d^*)-\sum_{i\in P_B, i<d^*}t(i)\cdot M_B(r(i), d^*),\] where $r(i)$ denotes the index of the pivot row corresponding to pivot column $i$. Since packet $d^*$ has been seen by A, $t(d^*)$ is uniformly distributed over $\F$ and is independent of other entries of $\mathbf{t}$. From this observation and the above expression for $t'(d^*)$, it follows that for any given $M_A$ and $M_B$, $t'(d^*)$ has a uniform distribution over $\F$, and the probability that it is not zero is therefore $\left(1-\frac 1q\right)$.
\endproof 

For the queuing analysis, we assume that a successful reception always causes the receiver to see its next unseen packet, provided the transmitter has already seen it. A consequence of this assumption is that the set of packets seen by a node is always a contiguous set, with no gaps in between. In particular, there is no repeated ACK due to packets being seen out of order. The above lemma argues that these assumptions become more and more valid as the field size increases. In reality, some packets may be seen out of order resulting in larger queue sizes. However, we believe that this effect is minor and can be neglected for a first order analysis.

\noindent {\it The expected queue size:}

We define arrival and departure as follows. A packet is said to arrive at a node when the node sees the packet for the first time. A packet is said to depart from the node when the node drops the witness of that packet from its queue. For each intermediate node, we now study the expected time between the arrival and departure of an arbitrary packet at that node. This is related to the expected queue size at that node, by Little's law. 

{\it Proof of Theorem \ref{thm:queue}:}

\IEEEproof
Consider the $k^{th}$ intermediate node, for $1\le k< N$. The time a packet spends in this node's queue can be divided into two parts:

1) {\it Time until the packet is seen by the sink:} 
	
	The difference between the number of packets seen by a node and the number of packets seen by the next node downstream essentially behaves like a $Geom/Geom/1$ queue. The Markov chain governing this evolution is identical to that of the virtual queues studied in \cite{ARQforNC}. Given that a node has seen a packet, the time it takes for the next node to see that packet corresponds to the waiting time in a virtual queue. For a load factor of $\rho$ and a channel ON probability of $\mu$, the expected waiting time was derived in \cite{ARQforNC} to be $\frac{(1-\mu)}{\mu(1-\rho)}$, using results from \cite{hunterbook}. Now, the expected time until the sink sees the packet is the sum of $(N-k)$ such terms, which gives $\sum_{i=k}^{N-1} \frac{(1-\mu_i)}{\mu(1-\rho_i)}$. 

 2) {\it Time until sink's ACK reaches intermediate node:} 
	
	The sink's ACK has to propagate from the source to the intermediate node in question through the feed-forward mechanism. Given that a node knows that the sink has seen the packet in question, the time it takes for the next node to get this information is the expected time until the next slot when the channel is ON. Since the $i^{th}$ channel is ON with probability $\mu_i$ in every slot, this expected time is simply $\frac 1{\mu_i}$. Thus, the time it takes for the sink's acknowledgment of the packet to propagate to node $k$ is given by $\sum_{i=1}^{k-1}\frac 1{\mu_i}$.

Thus, the total expected time a packet spends in the queue at the $k^{th}$ node ($1\le k <N$) is given by:
\[\mathbb{E}[T_k]=\sum_{i=k}^{N-1} \frac{(1-\mu_i)}{\mu_i(1-\rho_i)}+\sum_{i=1}^{k-1}\frac 1{\mu_i}\]
Assuming the system is stable (\ie, $\lambda<\min_i\mu_i$), we can use Little's law to derive the expected queue size at the $k^{th}$ node:
\[\mathbb{E}[Q_k]=\sum_{i=k}^{N-1} \frac{\rho_i(1-\mu_i)}{(1-\rho_i)}+\sum_{i=1}^{k-1}\rho_i\]
\endproof

\section{Conclusions and future work}\label{sec:conc}
In this work, we propose a new approach to congestion control on lossy links based on the idea of random linear network coding. We introduce a new acknowledgment mechanism that plays a key role in incorporating coding into the control algorithm. From an implementation perspective, we introduce a new network coding layer between the transport and network layers on both the source and receiver sides. Thus, our changes can be easily deployed in an existing system. A salient feature of our proposal is that coding operations occur only at the end hosts, thereby preserving the end-to-end philosophy of TCP. 

We observe through simulations that the proposed changes lead to huge throughput gains over TCP in lossy links. For instance, in a 4-hop tandem network with a 5\% loss rate on each link, the throughput goes up from about 0.007 Mbps to about 0.39 Mbps for the correct redundancy factor.

In the future, we plan to understand the impact of field size on throughput. While our current simulations assume a large field size, we believe that in practice, a large part of the gains can be realized without too much overhead. We also wish to understand the overhead associated with the coding operations in a practical setting. Throughput gains are seen even though the intermediate nodes do not perform any coding. Theory suggests that a lot can be gained by allowing intermediate nodes to code as well. Quantifying the impact of such coding is of interest in the future. 

This paper presents a new framework for combining coding with feedback based rate-control mechanisms in a practical way. It is of interest to extend this approach to more general settings such as network coding based multicast over a general network. Even in the point-to-point case, we could use these ideas to implement a multipath-TCP based on network coding.
\bibliographystyle{IEEEtran}
\bibliography{InfocomReferences}
\end{document}

%% file: preamble.tex
\usepackage{epsfig}
\usepackage{graphicx}
\usepackage{amsmath}
\usepackage{times}
\usepackage{amssymb}
\usepackage{amsfonts}
\usepackage{hyperref}

\newtheorem{theorem}{Theorem}
\newtheorem{definition}{Definition}
\newtheorem{lemma}{Lemma}
\newtheorem{proposition}{Proposition}
\newtheorem{algorithm}{Algorithm}[section]
\newtheorem{claim}{Claim}[section]
\newtheorem{corollary}{Corollary}

\newtheorem{remark}{Remark}

\newcommand{\F}{\mathbb{F}_q}

\newcommand{\beqn}{\begin{equation}}
\newcommand{\eeqn}{\end{equation}}

\newcommand{\ie}{{\it i.e.}} 
\newcommand{\etal}{\textit{et al. }}